# A Video-Aware FEC-Based Unequal Loss Protection System for Video Streaming over RTP

César Díaz, Julián Cabrera, Fernando Jaureguizar and Narciso García

*Abstract* — *A video-aware unequal loss protection (ULP) system for protecting RTP video streaming in bursty packet loss networks is proposed. Considering the relevance of the frame, the state of the channel, and the bitrate constraints of the protection bitstream, our algorithm selects in real time the most suitable frames to be protected through forward error protection (FEC) techniques. It benefits from a wise RTP encapsulation that allows working at a frame level without requiring any further process than that of parsing RTP headers. This makes our system straightforward and fast, perfectly suitable to be included in commercial video streaming servers. Simulation results show how our technique outperforms other proposed ULP schemes[1].*

## I. INTRODUCTION

Internet real-time video streaming applications have gained much importance over the last years. Hence, dealing with the lossy nature of IP-based networks in time-sensitive scenarios has become a crucial task. In that sense, FEC-based data protection schemes constitute the most suitable choice in many real-time environments, as no extra delay is added due to retransmissions [1]. Moreover, as resources might be limited, smart strategies are usually introduced to decide which part of the data should be protected and how, so that resource availability is not exceeded and the overall quality after decoding is kept as high as possible. These are called unequal loss protection (ULP) schemes [2].

Different ULP techniques have been proposed in the literature. They usually differ in two main aspects which influence the computational cost of the scheme: the scope of the decisions taken and the level at which the analysis on video data is performed. The first aspect refers to the structure of data over which a decision is taken: a set of packets in a stream, a set of macroblocks in a frame, a frame, a video layer… The second one alludes to the units within the encoded video stream whose features are analyzed to perform the prioritization: macroblock ranking, frame classification, video scalability exploitation… [3]-[6]. In general, the finer the granularity of evaluation is, the more computationally costly the technique is. Once data are accessed and analyzed, most of the techniques raise cost minimization problems whose solutions determine the behavior of the scheme, that is, the protection policies to follow. The cost function to be minimized is typically based on a model of the distortion that affects the video when a portion of the information is lost [7], [8].

Most of the proposed ULP schemes are effective at distortion minimization. However, due to the strong restrictions imposed by real-time applications, it is also necessary for a scheme that it can be carried out efficiently. In that sense, the video data processed by the algorithm must be easily and quickly obtained and analyzed. For that purpose, both the scope of decision and the scope of analysis need to be adapted to fit those requirements.

In that sense, we propose a smart ULP scheme, Video-Aware ULP (VA-ULP) scheme, which works from both perspectives (decision and analysis) at a frame level. Taking advantage of a previous wise RTP encapsulation of the encoded video stream [9], working at a frame level does not require to our algorithm any further process than that of parsing RTP headers. So the access to the required data and their evaluation is straightforward and fast.

The proposed scheme first selects in real time the most suitable frames to be protected. Then it applies a predetermined FEC code to those RTP packets that wrap the selected frames. Decisions are taken considering: (i) the type of frame and the GOP structure of the video stream in terms of minimizing the error propagation (and therefore the decoded video distortion); (ii) the behavior of the transmission channel through an appropriate model; and (iii) the limited resources for the protection stream.

The VA-ULP scheme is presented as part of a modular protection unit, included in the video server, in which every submodule fulfills a specific task derived from the global one: the smart generation of a protection stream. Those tasks include: RTP rewrapping, data access and evaluation, execution of the VA-ULP algorithm, and creation of the FEC packets.

The rest of the paper is organized as follows. In section II, the main actors of the video streaming system are described. Section III introduces the different blocks and submodules that compose the protection module, jointly with the VA-ULP scheme. In section IV, the VA-ULP algorithm is explained in depth. Simulations and results are presented in section V.

[1] This work has been partially supported by Alcatel-Lucent and the Spanish Administration agency CDTI under project CENIT-VISION 2007-1007, by the Ministerio de Ciencia e Innovación of the Spanish Government under project TEC2010-20412 (Enhanced 3DTV), and by the Ministerio de Industria under project ITEA2-JEDI.

César Díaz, Julián Cabrera, Fernando Jaureguizar and Narciso García are with the Grupo de Tratamiento de Imágenes (GTI), Universidad Politécnica de Madrid, 28040 Madrid, Spain (e-mail: {cdm, julian.cabrera, fjn, narciso}@gti.ssr.upm.es).

Finally, in section VI we include the conclusions of the paper.

## II. SYSTEM MAIN ACTORS

The main actors in any video streaming system are the encoded video stream and the transmission channel. Therefore, the problem of selecting the most suitable data to be protected is formulated through the description of their behavior. This characterization is presented next.

### A. Encoded video stream

The encoded video stream is simply characterized as a sequence of frames presenting different features in relation to error propagation. These features are: frame type (I, P or B), frame size, and distance to the end of the GOP. The frame type directly specifies the hierarchy coding level of the frame. The frame size is relevant since it directly depends on the amount of information given by that frame. The distance to the end of the GOP indicates the likelihood that the frame could be referred by a large number of other frames within the GOP [10].

### B. Transmission channel

In IP-based networks, packet losses are of a bursty nature. These bursts can be described in terms of frequency of occurrence and length. That characterization can lead to the generation of models from which it is possible to predict the behavior of the channel at a packet level to some extent.

One of the most used models is the simplified Gilbert-Elliot model [11], illustrated in Fig. 1.

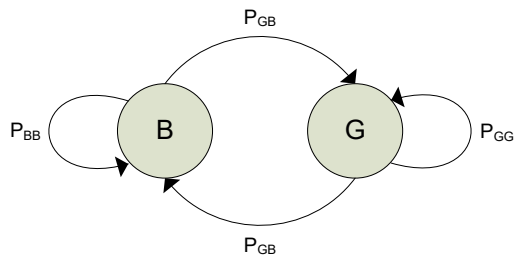

**Fig. 1 Simplified Gilbert-Elliot model**

State $G$ indicates that the transmitted packet has been successfully received and state $B$ indicates that the transmitted packet has been lost in the channel. The parameters $P_{GG}$, $P_{GB}$, $P_{BG}$ and $P_{BB}$ represent the transition probabilities and $P_G$ and $P_B$ are the stationary probability of being in state $G$ and in state $B$, respectively. Their values can be calculated as a function of the Packet Loss Rate (PLR) and the Average Burst Length of the channel (ABL).

## III. PROTECTION MODULE

The protection unit is presented in Fig. 2. It is divided into two main blocks: a first block, the rewrapper, that receives, analyzes and rebuilds the encoded video stream, performing the RTP reencapsulation; and a second one, called VA-ULP module, which is in charge of analyzing both the video stream and the channel behavior, executing the VA-ULP algorithm, and generating and streaming the protection packets.

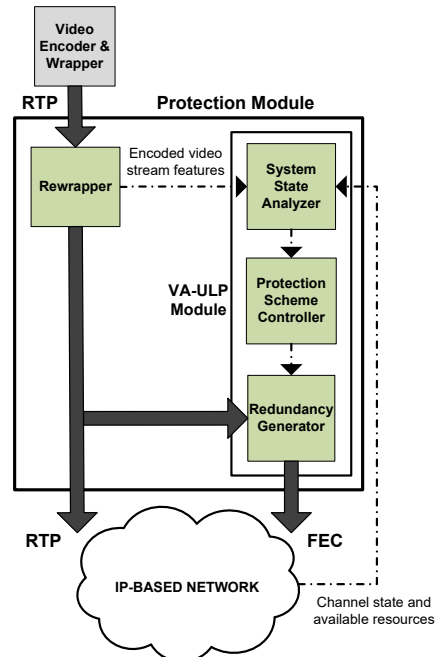

**Fig. 2 Block diagram of the video server, including the protection module**

These two blocks are described more in detail in the following subsections.

### A. Rewrapper

The input to the protection system is an RTP stream encapsulating MPEG2-TS high quality video data. Through the rewrapper [9], TS data packets are regrouped, labeled and reencapsulated in RTP packets, so that each RTP packet contains data of a single video frame and all data packets corresponding to a frame are streamed sequentially. Moreover, an RTP header extension has been created with flags notifying the type of the information carried, the beginning and end of the frame and the distance to the end of the GOP (for video data).

### B. VA-ULP module

The problem of selecting the more convenient frames to be protected is solved through our VA-ULP scheme. It selects the most suitable protection policy, i.e., the one that, fulfilling a certain imposed bitrate constraint, points out which video frames among the whole set should be protected to minimize a certain cost function. Decisions are thus reached at a frame level: the whole set of RTP packets wrapping information of a frame are either protected or not protected.

Optimal minimum distortion results could only be achieved if all the video frames in a sequence were considered together in the optimization problem. However, the high computational burden required for this approach together with real-time constraints forces to work with subsets of consecutive frames, leading to suboptimal but realizable approaches. We call these sets of pictures Decision Frame Sets (DFSs). The internal structure of a DFS is shown in Fig. 3.

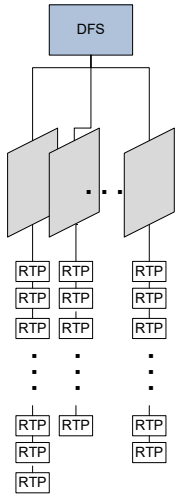

**Fig. 3 Internal structure of a Decision Frame Set**

The number of frames that a DFS contains is a trade-off between the accuracy of the protection approach (the higher the number of frames in a DFS is, the better the optimization approach can be) and the extra latency that is added as the number of frames increases.

As the loss of I-frame data may lead to major error propagation within the whole GOP, the first and main aim of our scheme is to protect the largest amount of this type of frames. For that purpose, and resorting to the basic distinction of frame type, DFSs are classified into sets containing an I-frame (I-DFSs) and sets containing no I-frames (only P- and B-frames) (PB-DFSs). The behavior of the scheme is different depending on the type of DFS.

The VA-ULP scheme assumes that there is a certain bitrate devoted for protecting the main stream. This bitrate is initially distributed equally among all the DFSs resulting in the nominal bit budget for the protection of each DFS. Nevertheless, it is very likely that I-DFSs require a greater budget than that of the nominal value. The reason is that I-frames, due to its compression characteristics, are of larger size than P- and B-frames. The VA-ULP scheme aims at preserving as much I-frames as possible in order to minimize error propagation. Thus, the protection system devotes a portion of the budget nominally assigned to PB-DFSs to the protection of I-DFSs.

The submodules that make up the VA-ULP module are described below.

*1) System State Analyzer*

This submodule is in charge of obtaining the VA-ULP scheme input data: the features of the video frames are fetched from the information contained in the RTP packet headers at the rewrapper output. The state of the transmission channel is acquired through a backward channel, whereas the budget for the protection of each DFS is updated after each decision is taken.

*2) Protection Scheme Controller*

Here, the VA-ULP algorithm is executed. For that purpose, the nominal budget and the portion of it devoted to the protection of I-DFSs are initially estimated. The algorithm solves a cost minimization problem in order to take a decision on every DFS.

*3) Redundancy Generator*

The protection packets corresponding to the frames selected by the scheme policy are generated in this submodule. The selected FEC technique is the 1-D interleaved parity FEC scheme proposed by the Pro-MPEG Forum in its COP #3 [13]. This code is based on the arrangement of consecutive data packets in matrices. FEC packets are generated column-wise. The reason to that choice is the simplicity and speed involved in XOR operation coding and its good performance in bursty loss channels, thanks to interleaving.

## IV. VA-ULP ALGORITHM

After the description of the protection module, the main aspects of the implementation are introduced here.

*A. Budget estimation*

- **Nominal budget calculation**

The nominal bit budget for the protection of each DFS, which is referred as $N_{bit\,FEC}$, is computed as shown in (1):

$$N_{bit\,FEC} = \left(R_{protection}(\text{bps})/R_{source}(\text{fps})\right)\cdot N_{frames\,DFS} \quad (1)$$

where $R_{protection}$ is the average bit-rate for protection purposes, $N_{frames\,DFS}$ is the number of frames per DFS and $R_{source}$ is the video framerate.

That budget can also be expressed as the number of protection packets that can be generated without exceeding the available resources, $N_{pkt\,FEC}$, just by dividing by the average protection packet length, $L_{pkt\,FEC}$. This is expressed in (2):

$$N_{pkt\,FEC} = N_{bit\,FEC}/L_{pkt\,FEC} \quad (2)$$

$L_{pkt\,FEC}$ can be empirically estimated from the average size of the first certain number of RTP data packets plus the bytes corresponding to the RTP-FEC header [12].

From the previous relation we can easily calculate the number of RTP data packets that can be nominally protected per DFS, $N_{pkt\,RTP}$, as presented in (3):

$$N_{pkt\,RTP} = (n/k)\cdot N_{pkt\,FEC} \quad (3)$$

where $n/k$ is the relation between the number of data packets plus the number of protection packets in a FEC Code ($n$) and the number of data packets to be protected ($k$). That means we are be able to protect $N_{pkt\,RTP}$ packets at every DFS.

- **To-be-reserved budget estimation**

From every PB-DFS, a portion of the nominal budget is to be reserved for forthcoming I-DFS' use. This value must cover most of the I-frames in the video stream. Given that the size of I-frames varies, we model it as a random variable and assume that it follows a Gaussian distribution. Hence, we can determine the minimum size threshold, $L_{\%}$, that guarantees that $p$ (%) of I-frames has a size of equal to or less than $L_{\%}$. That is illustrated in Fig. 4.

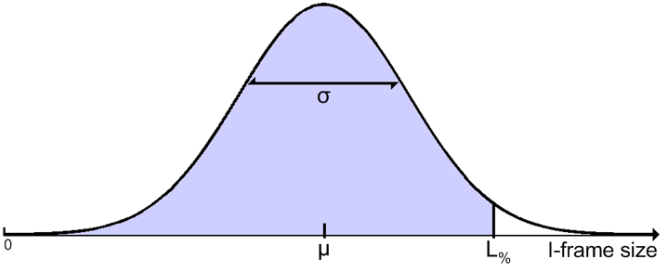

Fig. 4 I-frame size Gaussian distribution

where $\mu$ is the average I-frame size and $\sigma$ is the standard deviation of the distribution. These parameters are estimated from studying the size of the I-frames along the main stream.

The size threshold, $L_{\%}$, is calculated as shown in (4):

$$L_{\%} = \mu + \sqrt{2} \cdot \sigma \cdot erf^{-1}(2 \cdot (p/100) - 1) \quad (4)$$

This way, to guarantee the protection of any I-frame of size $L_{\%}$ or less, belonging to the forthcoming I-DFS, it is necessary to reserve $N_{pkt\ RTP\ threshold}$ packets from every PB-DFS. $N_{pkt\ RTP\ threshold}$ is dependent on the threshold, $L_{\%}$, on the average GOP length $L_{GOP}$ and on the number of frames per DFS, $N_{frames\ DFS}$, and is calculated as (5):

$$N_{pkt\ RTP\ threshold} = (L_{GOP} / N_{frames\ DFS}) \cdot L_{\%} \quad (5)$$

On the other hand, if the nominal budget for the protection of the DFSs is inferior to $N_{pkt\ RTP\ threshold}$, the whole budget is then reserved. Thus, finally, the number of packets reserved at every PB-DFS, $N_{pkt\ RTP\ reserved}$, is expressed in (6):

$$N_{pkt\ RTP\ reserved} = \min(N_{pkt\ RTP\ threshold}, N_{pkt\ RTP}) \quad (6)$$

### B. Algorithm description

For every DFS, the VA-ULP algorithm performs as follows:

- **In the case of PB-DFS**: $N_{pkt\ RTP\ reserved}$ packets of the nominal budget are reserved for the protection of the forthcoming I-DFS. The non-reserved portion of the nominal budge is used for the protection of the frames contained in this PB-DFS. A cost minimization problem is then solved to decide which frames should be protected.
- **In the case of I-DFS**: no bits are reserved. The whole nominal bit budget for the current DFS and the extra resources reserved from previous PB-DFSs are used for the protection of this I-DFS. Again, a cost minimization problem is raised.

The different aspects of the cost minimization problem are introduced below.

#### 1) Cost function

At the $k^{th}$ DFS, the cost to be minimized, $c_k$, is equal to the sum of the expected distortion introduced by each of the $N_{frames}$ frames belonging to that DFS.

The expected distortion introduced by a frame represents the potential quality degradation of the decoded video stream due to the possible loss of information in the transmission of that frame. It depends not only on the specific coding features of the frame (frame type, distance to the end of the GOP, and frame size), but also on the behavior of the channel and on its initial state, and on the decision taken over that frame.

The coding features of the frames in a DFS are represented through $N_{frames}$-component vectors, where each component refers to a frame. We define $t_k$, $d_k$, and $z_k$ as coding feature vectors that reflect the type of frame, the distance to the end of the GOP, and the size of each frame in the $k^{th}$ DFS, respectively. These vectors are grouped in the DFS state matrix, $x_k$, which is made up of three rows, corresponding to the three coding features vectors, and $N_{frames}$ columns. Therefore, the $i^{th}$ column of $x_k$ represents the coding features of the $i^{th}$ frame as presented in (7):

$$x_k(i) = (t_k(i), d_k(i), z_k(i))^T \quad (7)$$

The initial state of the transmission channel is introduced by the variable $s_k$. The decision taken over the frame is represented by the $N_{frames}$-component vector $\pi_k$. The component $\pi_k(i)$ is equal to 1 if the $i^{th}$ frame is decided to be protected, and equal to 0 otherwise.

At the $k^{th}$ DFS, the cost function is therefore expressed as the sum of the expected distortions of all the frames in the DFS, as shown in (8):

$$c_k(x_k, s_k, \pi_k) = \sum_{i=1}^{N_{frames}} D_i(x_k(i), s_k, \pi_k(i)) \quad (8)$$

where $D_i$ is the expected distortion of the $i^{th}$ frame.

We assume that the distortion introduced by a frame only depends on whether it has been completely received or not. Therefore, we consider that when a single packet cannot be recovered, this leads to the loss of the whole frame. Thus, the expected distortion of a given frame, $D_i$, is expressed as presented in (9):

$$D_i(x_k(i), s_k, \pi_k(i)) = D_{iR}(x_k(i)) \cdot P_R(s_k, \pi_k(i)) \\ + D_{iW}(x_k(i)) \cdot P_W(s_k, \pi_k(i)) \quad (9)$$

where $D_{iR}$ is the frame distortion when all the packets of the frame are recovered, $D_{iW}$ is the frame distortion when at least one of the packets cannot be recovered, and $P_R$ and $P_W$ are, respectively, the likelihood of these two complementary events.

$D_{iR}$ is related to the video coding process and is only dependent on the applied quantizer and other coding parameters. As our protection system does not get involved in this process and, moreover, coding distortion is far lower than the one introduced when the frame is lost, it is assumed that $D_{iR} = 0$. So, the expected distortion is presented in (10):

$$D_i(x_k(i), s_k, \pi_k(i)) = D_{iW}(x_k(i)) \cdot P_W(s_k, \pi_k(i)) \quad (10)$$

The probability of losing at least one packet of a frame, $P_W$, is higher for non-protected frames than for protected ones. Therefore, for a given frame, the expected distortion is higher for non-protected frames, as expressed in (11):

$$D_i(x_k(i), s_k, \pi_k(i) = 1) < D_i(x_k(i), s_k, \pi_k(i) = 0) \quad (11)$$

The goal of the problem is to select for each DFS the control policy that minimizes the cost, i.e., the global expected distortion introduced by the frames of the DFS, given a limitation in the available bitrate for data protection.

The two terms on which the expected distortion depends (the distortion when the frame is lost, $D_{iW}$, and the likelihood of losing information of a frame, $P_W$) are described next.

*2) Distortion model*

The distortion model, $D_{iW}$, does not depend on actual PSNR measurements but on frame coding features. So it considers that the loss of packets of a certain frame leads to a quality drop that directly depends on the importance of that frame, i.e., on its coding features: the more relevant the video frame is, the greater the distortion becomes if that frame is lost. That helps our system be straightforward and fast, as no actual PSNR measurements are needed during the execution of the algorithm. $D_{iW}$ is modeled as dependent on the type of video frame, $t_k(i)$, the distance to the end of the GOP, $d_k(i)$, and the size of the frame, $z_k(i)$, through the equation presented in (12):

$$\begin{aligned} D_{iW}(x_k(i)) &= D_{iW}(t_k(i), d_k(i), z_k(i)) = \\ &\quad K_1(t_k(i)) + K_2 \cdot d_k(i) + K_3 \cdot z_k(i) \end{aligned} \quad (12)$$

where $K_1$ is a constant that directly depends on the type of video frame, and $K_2$ and $K_3$ are weighting factors for the size of the frame and the distance to the end of the GOP, respectively. These factors have been tuned empirically after the analysis of numerous video transmissions.

*3) Likelihood of losing information of a frame*

The likelihood of not being able to recover one or more data packets of a given video frame is expressed in terms of the transmission channel model. It depends on whether this frame is protected or not. So, $P_W(s_k, \pi_k(i)=0)$ represents the likelihood of losing at least one packet of a frame when this frame is not protected and $P_W(s_k, \pi_k(i)=1)$ represents the likelihood of not being able to recover all the packets of a frame when this frame is protected.

- **Non-protected frame**

For determining $P_W(s_k, \pi_k(i)=0)$ the two possible initial channel states are contemplated: the case of having successfully received the last packet (coming from state $G$) and the case of having lost it (coming from state $B$). So, $P_W(s_k=G, \pi_k(i)=0)$ indicates the likelihood of losing at least one packet of a frame after the previous packet was successfully received and $P_W(s_k=B, \pi_k(i)=0)$ is the likelihood of that event after the previous packet was not received. The likelihood of both events can be computed from the transition probabilities of the channel mode, as presented in (13) and (14):

$$P_W(s_k = G, \pi_k(i) = 0) = 1 - P_{GG}^{z_k(i)} \quad (13)$$

$$P_W(s_k = B, \pi_k(i) = 0) = 1 - P_{BG} \cdot P_{GG}^{(z_k(i)-1)} \quad (14)$$

- **Protected frame**

The computation of $P_W(s_k, \pi_k(i)=1)$ is more complex since we have to consider the recovery capabilities of the FEC protection scheme.

Since we are applying the 1-D interleaved parity FEC technique, protection packets are generated from sets of $D \cdot L$ consecutive RTP packets, where $D$ is the number of rows and $L$ the number of columns of the protection matrix. Therefore, given a choice for $D$ and $L$, $N_{matrices}$ protection matrices are necessary to protect a video frame of size $z_k(i)$, as it is expressed in (15):

$$N_{matrices} = \lceil z_k(i)/(D \cdot L) \rceil \quad (15)$$

Therefore, $P_W(s_k, \pi_k(i)=1)$ can be expressed as the likelihood of not being able to recover one or more of the data packets of any of the $N_{matrices}$ protection matrices. We assume that the likelihood of not being able to recover one or more data packets of a single protection matrix, $P_{W\,matrix}(s_k)$, is independent of that of the rest of the protection matrices, as reconstruction is performed independently for each matrix. Also assuming that $P_{W\,matrix}(s_k)$ is low enough, we can approximate:

$$\begin{aligned} P_W(s_k, \pi_k(i) = 1) &= 1 - (1 - P_{W\,matrix}(s_k))^{N_{matrices}} \\ &\approx N_{matrices} \cdot P_{W\,matrix}(s_k) \end{aligned} \quad (16)$$

We consider a realization of a given protection matrix as the result of the transmission of all its packets. Therefore, $P_{W\,matrix}(s_k)$ can be computed by adding the likelihood of all those matrix realizations in which all the frame packets cannot be recovered.

Due to the large number of different possible realizations, the direct computation of their probability can be a very time consuming process. Thus, a simplified method is proposed. In this simplified method it is assumed that: (i) all lost packets occur in bursts; and (ii) just one single burst takes place within a certain matrix. Therefore it is possible to approximate $P_{W\,matrix}(s_k)$ as the likelihood that a burst exceeding the correction capacity of the code occurs, i.e., that the length of the burst is greater than the number of columns ($L$) of the 1-D interleaved parity FEC matrix. That is expressed in (17):

$$P_{W\,matrix}(s_k) = \sum_{N=L+1}^{D \cdot L} P_{burst}(s_k, N) \qquad (17)$$

where $P_{W\,burst}(s_k, N)$ is the likelihood of occurrence of a burst of a length of $N$ packets. Its value can be calculated from the transition probabilities of the channel mode and the parameters of the FEC code.

Taking into consideration the two possible initial states, the likelihood of occurrence of a burst of a length of $N$ packets after the previous packet was successfully received is presented in (18):

$$P_{burst}(s_k = G, N) = \\ P_{GB} \cdot P_{BB}^{N-1} \cdot P_{GG}^{(D \cdot L - N - 1)} \cdot \left( (D \cdot L - N) \cdot P_{BG} + P_{GG} \right) \qquad (18)$$

On the other hand, the likelihood of occurrence of a burst of a length of $N$ packets after the previous packet was lost is expressed in (19) and (20):

$$P_{burst}(s_k = B, N < D \cdot L) = \\ P_{BG} \cdot P_{BB}^{N-1} \cdot P_{GG}^{(D \cdot L - N - 2)} \cdot \\ \left( (D \cdot L - N - 1) \cdot P_{BG} \cdot P_{GB} + P_{BB} \cdot P_{GG} + P_{GB} \cdot P_{GG} \right) \qquad (19)$$

$$P_{burst}(s_k = B, N = D \cdot L) = P_{BB}^{D \cdot L} \qquad (20)$$

## V. SIMULATIONS AND RESULTS

We compare our proposal to the following two strategies:

- **Uniform Protection scheme (UP)**: data packets are protected as long as the protection bitrate constraint is fulfilled. No prioritization is performed.
- **ULP scheme proposed in [4] (MP)**: low-complexity strategy that, as ours, works at a frame level and performs the data prioritization in function of the frame type and the distance to the end of the GOP. As long as the imposed protection bitrate is fulfilled, frames are protected following the next order: first I-frames, next P-frames from the farthest to the closest to the end of the GOP, and last B-frames from the farthest to the closest to the end of the GOP. Whole frames are either protected or not protected and the same FEC technique with the same parameters is applied to all the ones decided to be protected.

The input to our system is a MPEG2-TS HD video movie with an average bitrate of approximately 12 Mbps. The encoded sequence is made up of reference frames (I- and P-frames) and non-reference frames (B-frames).

The packet transmission channel is simulated through a simplified Gilbert-Elliot model [11] whose parameters were obtained by means of numerous video transmission tests in ADSL and 802.11 distribution networks in a home environment. These parameters are:
- ADSL channel → PLR ≃ 1%, ABL ≃ 10
- 802.11 channel → PLR ≃ 2%, ABL ≃ 20

As the transmission channel is simulated, its state can be simply acquired by checking the corresponding channel state variable.

As indicated, the applied FEC technique is 1-D interleaved XOR. The number of rows of the protection matrices, $D$, has been set to 4, and the number of columns, $L$, to 20 for both network channels.

The number of frames per DFS, $N_{frames\,DFS}$, has been set to 5, as empirical tests proved it to be a good trade-off. This value is used for both the VA-ULP and the MP schemes.

All these parameters are used for the calculations involved in the cost minimization problems formulated during the performance of our strategy.

Simulation results for packet loss recovery rate are presented in Fig. 5 and Fig. 6 for an ADSL channel and for an 802.11 channel, respectively. They show the percentage of packets of each type that the system has been able to recover using the different schemes and imposing different redundancy rates: 5%, 10%, 15%, and 20% of the main stream's bitrate.

The experimental results in terms of packet loss recovery rate first show that since the protection bitrate imposed to the three strategies is the same, their overall recovery rate approximately matches. However, the specific recovery rate for each packet type differs.

With respect to the UP scheme, it distributes uniformly the redundancy bitrate among data packets regardless of their type. Therefore, the type of packets that is more numerous (packets of type B) will be the most likely to be used in the FEC generation, thus the most likely to be recovered in case of losses.

Regarding the performance of the ULP schemes, as the protection of packets wrapping information of reference pictures (I- and P-frames) is strengthened, the amount of packets of this kind recovered increases with respect to the non-smart strategy.

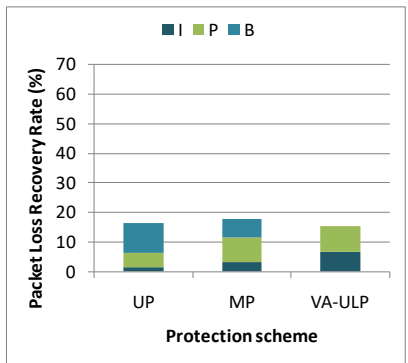
(a) Redundancy Rate = 5%

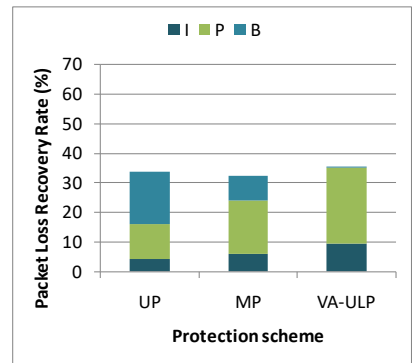
(b) Redundancy Rate = 10%

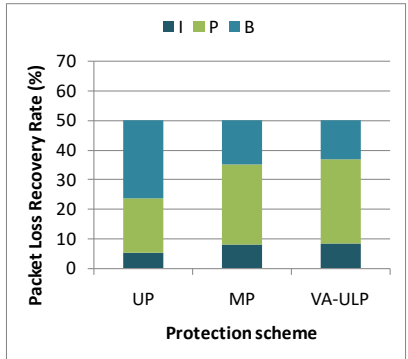
(c) Redundancy Rate = 15%

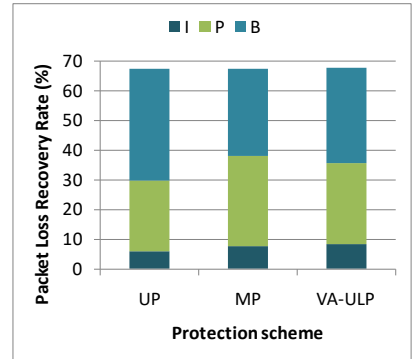
(d) Redundancy Rate = 20%

Fig. 5 Packet Loss Recovery Rate in an ADSL network channel

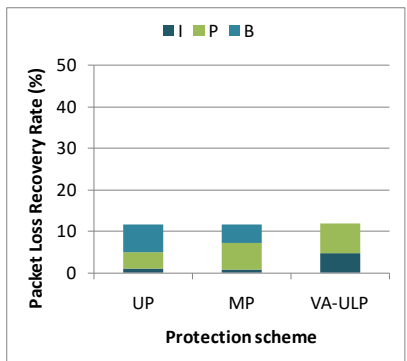
(a) Redundancy Rate = 5%

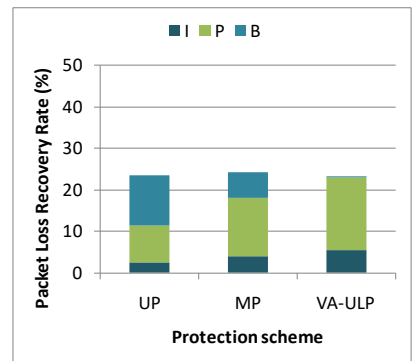
(b) Redundancy Rate = 10%

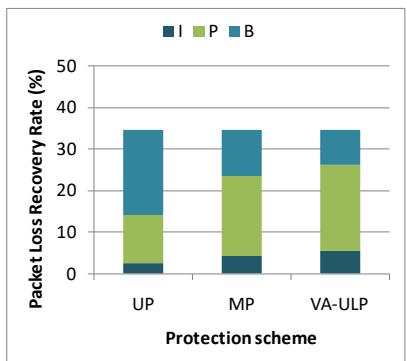
(c) Redundancy Rate = 15%

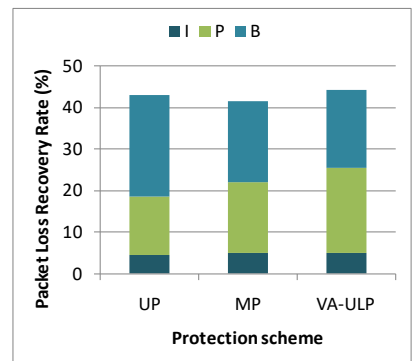
(d) Redundancy Rate = 20%

Fig. 6 Packet Loss Recovery Rate in an 802.11 wireless network channel

Nevertheless, the VA-ULP scheme manages to devote the protection rate mainly to I- and P-frames, in contrast to the MP algorithm, which also includes a significant number of B-frames. The inclusion of the budget reservation policy and the proposed distortion model allows the VA-ULP scheme to recover more packets of types I and P than the MP strategy at expenses of losing packets of type B. This behavior is more evident at lower protection bitrates.

The simulation results for average PSNR are provided in Fig. 7 and Fig. 8. It can be observed that the VA-ULP scheme outperforms both the UP and the MP schemes. This result validates, in terms of objective quality, both the budget reservation policy and the proposed distortion model, which not only takes into account the frame type and the distance to the end of the GOP, but also incorporates the channel behavior and the frame length.

As the redundancy rate is increased, the budget reservation policy influences less the obtained results, as every DFS counts on a larger nominal budget. Therefore, the PSNR gap of the VA-ULP scheme with respect to the MP scheme mostly relies on the distortion model.

Results for simulations in which no redundancy was used are also provided for comparison.

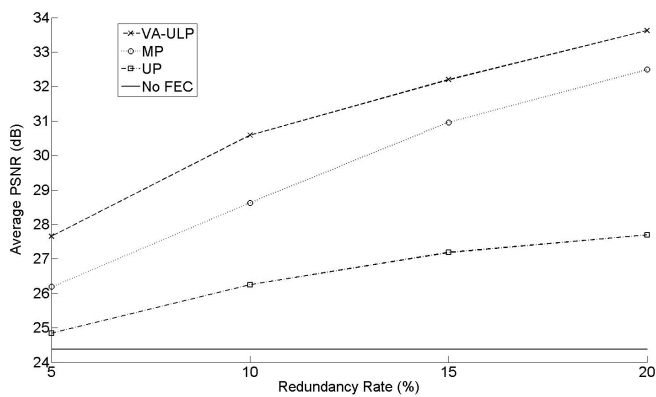

**Fig. 7 Average PSNR in an ADSL network channel**

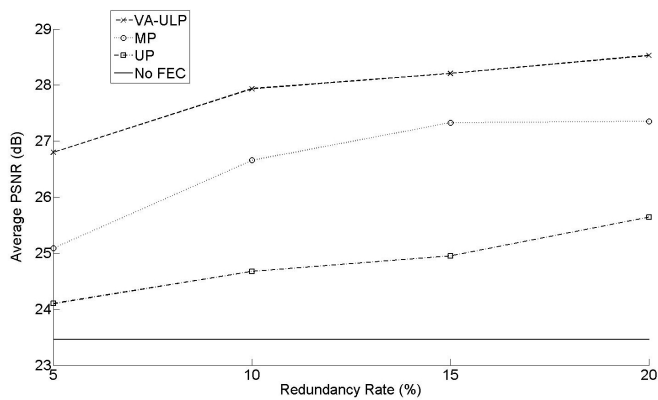

**Fig. 8 Average PSNR in an 802.11 wireless network channel**

## VI. CONCLUSIONS

We have designed a smart video-aware protection scheme for protecting RTP video streaming in bursty packet loss IP-based networks. It considers the relevance of the frames, the behavior of the channel, and the bitrate devoted to protection purposes to select in real time the most suitable parts of the main stream to be protected through FEC techniques. It works from the point of view of both decision and analysis at a frame level. Thanks to a wise RTP reencapsulation of the video stream, working at that level does not require any further process than parsing RTP headers.

The VA-ULP scheme is part of a modular protection unit, included in the video streaming server, whose aim is the smart generation of a secondary stream for protecting the main one.

A series of experiments have been carried out to compare our proposal with a non-smart strategy and the scheme proposed in [4]. Simulation results show how the proposed scheme outperforms both strategies thanks to a better management of the use of the limited protection bitrate.

**BIOGRAPHIES**

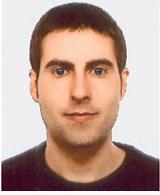
**César Díaz** received the Telecommunication Engineering degree (five-year engineering program) in 2007 and the Communication Technologies and Systems Master degree (two-year MS program) in 2010, both from the Universidad Politécnica de Madrid (UPM), Madrid, Spain. He is currently a PhD student at the same University.

Since 2008 he has been a member of the Grupo de Tratamiento de Imágenes (Image Processing Group) of the UPM. His research interests are in the area of video streaming/video transmission protection.

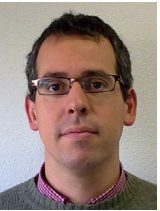
**Julián Cabrera** received the Telecommunication Engineering degree and the Ph.D. degree in Telecommunication, both from the Universidad Politécnica de Madrid (UPM), in 1996 and 2003, respectively.

Since 1996 he is a member of the Image Processing Group of the UPM. He was a Ph.D. scholar of the Information Technology and Telecommunication Programs of the Spanish National Research Plan from 1996 till 2001. Since 2001 he is a member of the faculty of the UPM, and since 2003 he is an Associate Professor of Signal Theory and Communications.

His professional interests include image and video coding, design and development of multimedia communications systems, focusing on Multivew Video Coding (MVC), 3D Video Coding and video transmission over variable rate channels. He has been actively involved in European projects (Acts, Telematics, IST) and national projects.

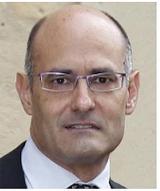
**Fernando Jaureguizar** received the Telecommunication Engineering degree and the Ph.D. degree in Telecommunication, both from the Universidad Politécnica de Madrid (UPM), in 1987 and 1994, respectively.

Since 1987 he is a member of the Image Processing Group of the UPM. He was a Ph.D. scholar of the Information Technology and Telecommunication Programs of the Spanish National Research Plan from 1988 till 1991. Since 1991 he is a member of the faculty of the UPM, and since 1995 he is an Associate Professor of Signal Theory and Communications.

His professional interests include digital image processing, video coding, 3DTV, and design and development of multimedia communications systems. He has been actively involved in European projects (Eureka, Acts, IST) and national projects.

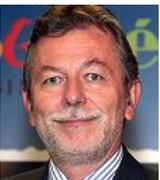
**Narciso García** received Telecommunication Engineering degree and the Ph.D. degree in Telecommunication, both from the Universidad Politécnica de Madrid (UPM), in 1976 (Spanish National Graduation Award) and 1983 (Doctoral Graduation Award), respectively.

Since 1977 he is a member of the faculty of the UPM, where is currently Professor of Signal Theory and Communications. He leads the Image Processing Group of the UPM. He was Coordinator of the Spanish Evaluation Agency from 1990 to 1992 and evaluator, reviewer, and auditor of European programs since 1990. His professional and research interests are in the areas of digital image and video compression and of computer vision.